\documentclass[%
 reprint,
showkeys,
 amsmath,amssymb,
 aip,pop,
]{revtex4-1}

\usepackage{graphicx}
\usepackage{dcolumn}
\usepackage{bm}
\usepackage{enumerate}
\usepackage{sidecap}

\def\apj{ApJ.}

\def\aap{A\&A}

\def\jgr{J. Geophys. Res.}
\def\grl{Geophys. Res. Lett.}
\def\apss{Astrophys. Space Sci.}
\def\apjl{Astrophys. J. Lett.}
\def\araa{Annual Review of Astronomy \& Astrophys.}

\def\sovast{Soviet Astronomy}
\def\cjaa{Chin. J. Astron. Astrophys.}
\def\pre{Phys. Rev. E}

\begin{document}

\preprint{APS/123-QED}

\title{Dynamic Topology and Flux Rope Evolution During Non-linear Tearing of 3D Null Point Current Sheets} 

\author{P. F. Wyper}
 \email{peterw@maths.dundee.ac.uk}
\author{D. I. Pontin}%
 \email{dpontin@maths.dundee.ac.uk}
\affiliation{%
 Division of Mathematics, University of Dundee,
    Dundee, UK} 


\date{\today}

\begin{abstract}
In this work the dynamic magnetic field within a tearing-unstable three-dimensional (3D) current sheet about a magnetic null point is described in detail. We focus on the evolution of the magnetic null points and flux ropes that are formed during the tearing process. Generally, we find that both magnetic structures are created prolifically within the layer and are non-trivially related. We examine how nulls are created and annihilated during bifurcation processes, and describe how they evolve within the current layer. The type of null bifurcation first observed is associated with the formation of pairs of flux ropes within the current layer. We also find that new nulls form within these flux ropes, both following internal reconnection and as adjacent flux ropes interact. The flux ropes exhibit a complex evolution, driven by a combination of ideal kinking and their interaction with the outflow jets from the main layer. The finite size of the unstable layer also allows us to consider the wider effects of flux rope generation. We find that the unstable current layer acts as a source of torsional MHD waves and dynamic braiding of magnetic fields. The implications of these results to several areas of heliophysics are discussed.
\end{abstract}

\keywords{Magnetic Reconnection, Magnetic Topology, MHD, MHD Waves}
\maketitle


\section{Introduction}
The topology of a magnetic field describes how its field lines are connected, and remains invariant if the field exists in a truly ideal plasma environment \cite{PriestForbes2000}. However, if the plasma is only close to ideal, the process of magnetic reconnection enables the magnetic topology to change -- liberating the free magnetic energy. Reconnection and topology change are central to many observed phenomena throughout the Heliosphere, including solar flares, geomagnetic storms in the Earth's magnetosphere and saw-tooth crashes in tokomaks, \cite[][and references therein]{Zweibel2009}.

Current sheets are a pre-requisite for the process of reconnection: within these structures the plasma can be sufficiently non-ideal that plasma and magnetic field become decoupled, allowing the magnetic connectivity to change. Understanding where current sheets form and how they behave is a crucial element of understanding reconnection, and consequently any phenomena that depend upon it. Important topological features common to astrophysical magnetic fields at which current sheets preferentially form include 3D magnetic null points -- isolated points in space at which the field strength is zero. In the solar atmosphere null points have been inferred to be abundant in the chromosphere and lower corona during quiet periods of the solar cycle \cite{Longcope2005,Regnier2008}, and during more active periods coronal null points are an predominant feature of active regions \cite{Zhao2008}. They have also been inferred to be involved in solar flares \cite{Masson2009,Aulanier2000,Fletcher2001}, jets \cite{Pariat2009,Pariat2010}, flux emergence \cite{Moreno-Insertis2008} and Coronal Mass Ejections (CMEs) \cite{Lynch2008,Antiochos1999}. 3D nulls have additionally been observed using {\it in situ} measurements from the Cluster satellites in the Earth's magnetotail \cite{Xiao2006} and are a prominent feature of the polar cusp regions \cite{Dorelli2007,Stern1973}. When combined with current sheets, null points are also excellent particle accelerators, \cite{Dalla2006,Stanier2012} and may be a contributing source of high energy particles in some solar flares \cite{Fletcher2001,Masson2009}.

It is well established that under the right conditions, current sheets will fragment via the tearing instability \citep{Furth1963}. Recently, it has been shown that even at Magnetohydrodynamic (MHD) scales, large aspect ratio current sheets are explosively unstable to this instability at the high magnetic Lundquist numbers typical of astrophysical plasmas \cite{Loureiro2007,Bhattacharjee2009}. When the field is two dimensional (2D), simulation studies have shown that the subsequent non-linear phase is dominated by magnetic island formation, coalescence and ejection, and that the average reconnection rate is only weakly dependent upon the magnetic dissipation \cite[e.g.][]{Huang2013}. 
However, when the field defining the current layer is fully three-dimensional these magnetic islands are replaced by flux ropes \cite[e.g.][]{Daughton2011,Markidis2013}. These helical regions of magnetic field are fundamental elements of evolving magnetic fields found at all scales throughout the Heliosphere; from laboratory experiments \cite{Intrator2009} to solar filaments, CMEs and interplanetary magnetic clouds \cite{Rust2003,Low2001,Lepping1990}. Therefore, understanding how flux ropes are generated and behave in the context of the reconnection process is also of major importance. 

\begin{SCfigure*}
\centering
\includegraphics[width=0.7\textwidth]{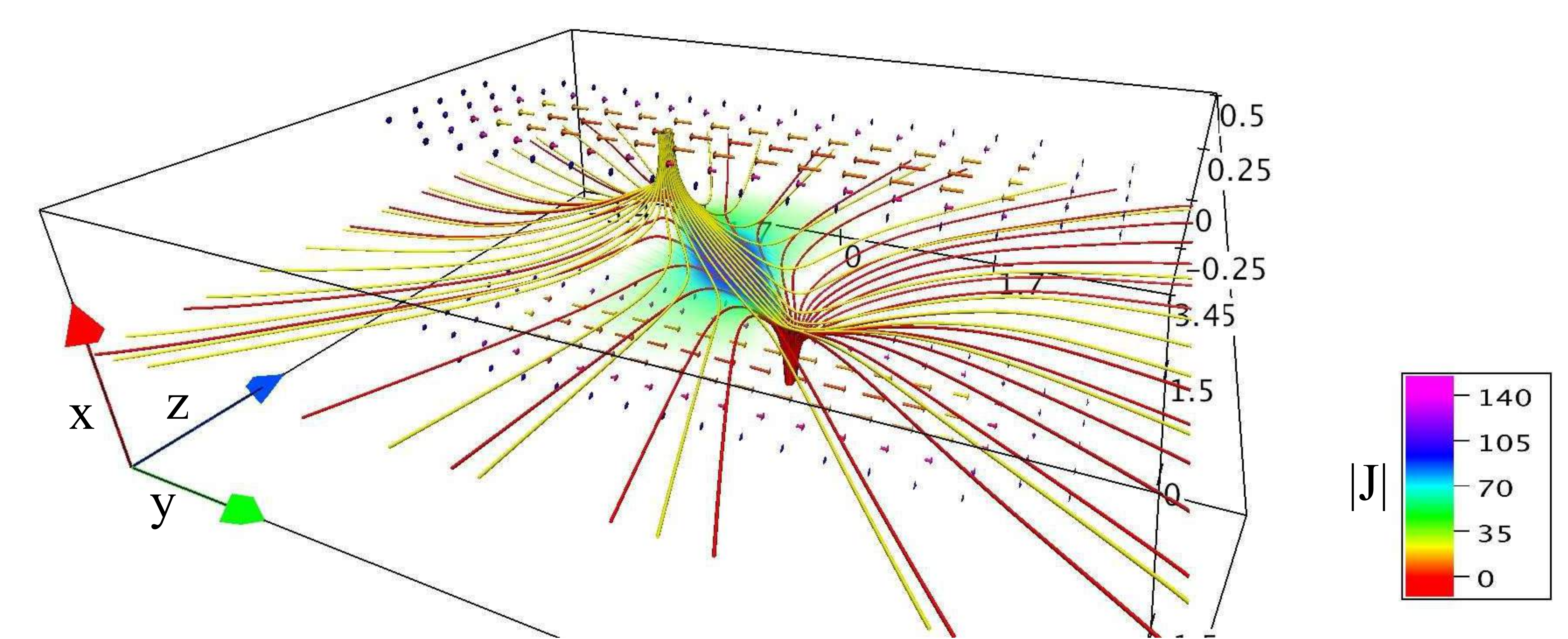}
\caption{(color online) The magnetic field in the simulation at $t=9$ (corresponding to the end of Stage I), depicting the setup of the experiment. Arrows show the tangential driving velocities applied on the $x$ boundaries, volume shading the current density and red and yellow field lines the 3D null point magnetic field structure. }
\label{fig:setup}
\end{SCfigure*}

In a recent series of numerical experiments we demonstrated that high aspect ratio current sheets formed at 3D null points fragment via the tearing instability, generating multiple evolving flux ropes which become heavily involved in the reconnection process, \citet{Wyper2014a} -- hereafter referred to as Paper 1. In addition, it was noted that multiple null points were formed during the process of fragmentation. Motivated by a desire to better understand how flux ropes, reconnection and topology change are interlinked, in this paper we focus on one particular numerical simulation and give a detailed account of the dynamics of the magnetic field following the onset of the tearing instability. In particular, we explain how and when new null points are created and annihilated, the formation and evolution of the flux ropes and how flux ropes and null points are coupled in this scenario. In Section II we describe the simulation setup, and in Section III summarise the evolution of the simulation. In Sections IV to VI the formation and annihilation of null points during the simulation is discussed alongside simple analytical models of the different bifurcations. Section VII then describes the evolution of the flux rope structures and their relation to 3D nulls. Finally, Sections VIII and IX discuss the significance of our results and summarise our findings.

\section{Simulation Setup}
The simulation was carried out using the Copenhagen staggered mesh code \citep{Galsgaard1997}. The simulation run that we focus on in this paper had a constant resistivity ($\eta$) value of $5\times 10^{-5}$, with a stretched grid of $[450,2000,200]$ points spread across a domain of $\pm[0.5,3.5,4]$ -- case 1 in Paper 1. The magnetic field in the volume contains a radially symmetric null at $t=0$, formed by placing two magnetic point sources outside the simulation box with strengths chosen so that in the vicinity of the null point the linearised field is given by $\mathbf{B} = [-2x,y,z]$. This initial equilibrium is disturbed by applying a tangential driving velocity on the $x$-boundaries, localised around the spine footpoints, see Fig.~\ref{fig:setup}. The driving has opposite sign at $x=\pm 0.5$, and is smoothly increased to a constant value of approximately $10\%$ of the local Alfv\'{e}n speed within one time unit, after which it remains constant. Length and time units are non-dimensionalised such that one unit of time is the Alfv\'{e}n travel time across one unit of length in a uniform plasma and magnetic field with $\rho=1$ and $|{\bf B}|=1$. The plasma is an ideal gas ($\gamma=5/3$), initially at rest with $e=0.025$ and $\rho=1$. All boundaries are closed and line-tied (${\bf B}\cdot{\bf n}$ fixed, ${\bf v}={\bf 0}$ outside driving regions). The mathematical expressions for the magnetic field and driver can be found in Paper 1.

\section{Stages of Evolution}
The evolution of the magnetic field passes through several phases. We first define each stage, giving a brief summary description, before considering the dynamics in detail further below.

\emph{I. Current Sheet Formation}: Once the driving begins, the footpoints of the spine lines are advected in opposite directions. A current layer forms in the weak field region around the null, generated by the local collapse of the spine and fan towards each other. As the driving continues the current sheet spreads across the fan surface, see Fig.~\ref{fig:setup}. 

\emph{II. Quasi-Steady Reconnection}: As the current intensity grows, spine-fan reconnection within the layer ensues, reconnecting fieldlines across the spines and the fan surface \cite{PontinBhatt2007}. The rate of reconnection becomes quasi-steady since the rate that flux is driven onto the layer is approximately balanced by the rate it is reconnected and ejected. The sheet continues to slowly lengthen and widen due to a slight imbalance of flux pile up at the edge of the current layer compared with the reconnection rate.

\emph{III. Primary Tearing}: Beyond a critical Lundquist number ($S_{c} \sim 2\times 10^{4}$, for details see Paper 1) the now high-aspect-ratio current sheet undergoes tearing, forming a symmetric flux rope pair that is ejected from the sheet by the out-flowing plasma. 

\emph{IV. Kinking Instability and Interaction}: Subsequent flux ropes form in the wake of the initial pair as the current layer becomes increasingly fragmented. With the symmetry of the sheet now broken these flux ropes are susceptible to a 3D instability \cite{Dahlburg2001,Dahlburg2002} that kinks them so that they interact and break up. At this stage the weak field region near the sheet center displays an increasingly turbulent field behavior, while further out  the layer is characterised by twisted writhing flux rope structures. 

Stages I and II have been investigated by a number of authors \cite[e.g.][]{PontinBhatt2007, Galsgaard2011, Wyper2012, Galsgaard2011b}. Following the identification of stages III and IV in Paper 1, we now aim to give a detailed account of the magnetic field evolution during these final two stages with the aim of better understanding the coupling between reconnection, flux rope formation and topology change.

\section{Null Formation}
The dynamics of the current layer in Stages III and IV is highly complex, with multiple flux rope and null point interactions. We begin by describing the evolution of the magnetic nulls, the predominant topological feature of our experiment. During Stages I and II the topology of the magnetic field remains relatively simple. The current sheet that forms cannot be a true discontinuity (due to the non-zero magnetic diffusion and finite resolution of the simulation grid), therefore the field contains a single, highly collapsed null point, i.e. a null with a very small angle between its spine and fan \citep{PontinBhatt2007}. Stage III is marked by the bifurcation of this null and the formation of helical field structures that we denote as ``flux ropes'' -- described in greater detail below. To observe this in our simulation we tracked the number and position of the magnetic nulls using the trilinear method, described by \citet{Haynes2007}. The magnetic structure in the vicinity of a generic 3D null point is given to first order by the linear terms of a Taylor expansion: $\mathbf{B}_{null} = \mathcal{M}\mathbf{x}^*$, where $\mathcal{M}$ is the Jacobian matrix evaluated at the null, $\mathbf{x}^* = [x-x_n,y-y_n,z-z_n]^T$, and the null is located at $(x_n,y_n,z_n)$ \cite[e.g.][]{Parnell1996}. The eigenvalues and eigenvectors of $\mathcal{M}$ at a given null dictate the topological degree (t.d.) of the null ($-1$ or $+1$) \cite{greene1993}, its nature (spiral or radial) as well as the orientation of the spine lines and fan surface. Sixth-order spatial derivatives (matching those from the numerical scheme) are used to accurately construct the Jacobian of the magnetic field for each null point.

We identify two predominant null point bifurcation processes occurring during the formation and ejection of flux ropes in our simulation. The first, denoted the primary bifurcation, occurs during the initial formation of the flux ropes following the tearing instability, and has a direct analogue with the change in topology in 2D tearing. Bifurcations of the second type, referred to hereafter as secondary bifurcations, occur as a result of internal reconnection within newly formed flux ropes, and have no direct analogue in 2D. In what follows we discuss both classes of bifurcation within the context of the first flux rope pair formation and ejection (Stage III), and present simple analytical models to describe them.

\begin{figure}
\centering
\includegraphics[width=0.5\textwidth]{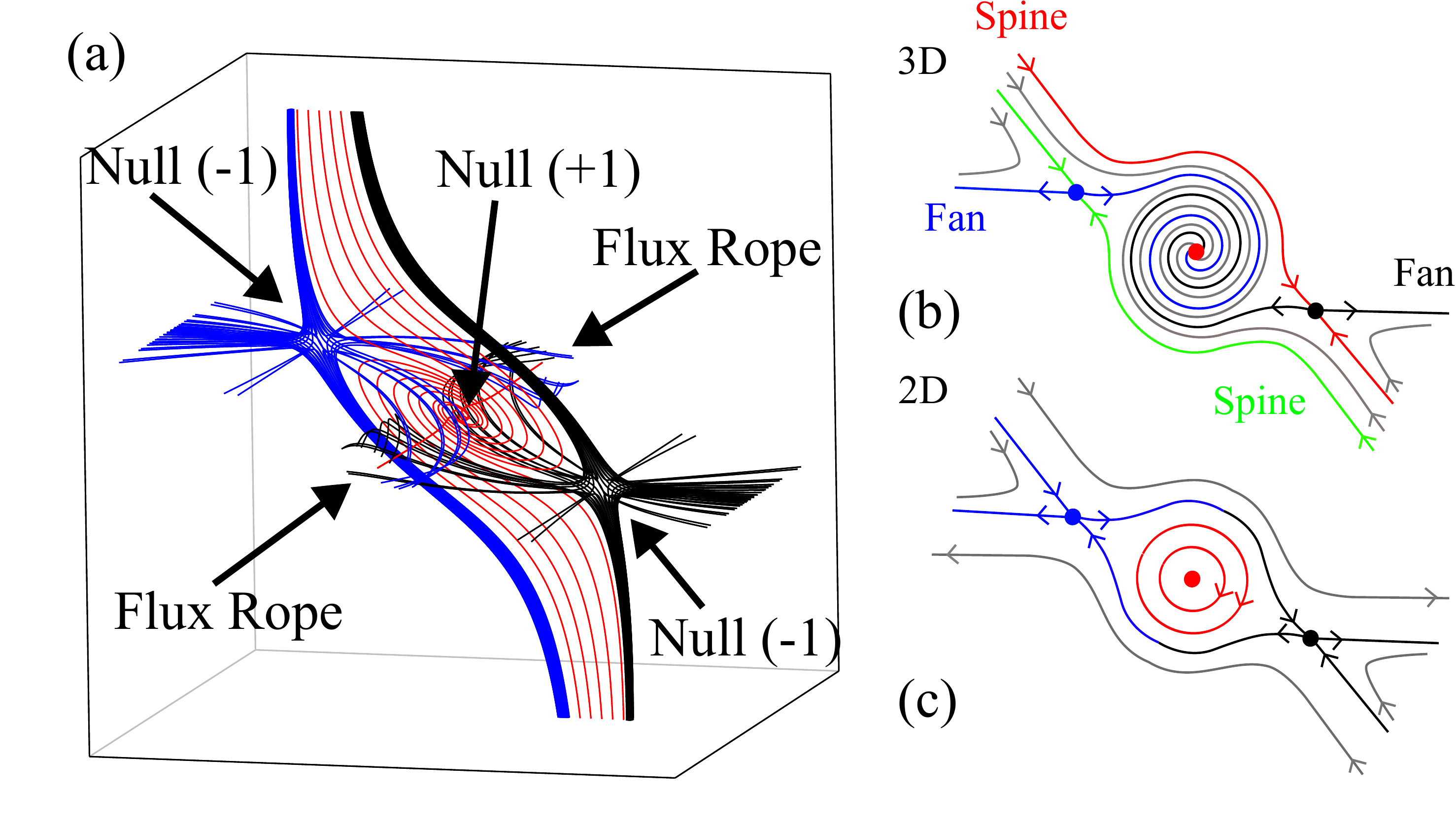}
\caption{(color online) Model of the field structure immediately after the primary bifurcation. (a) the 3D field structure; (b) schematic of the field in the $xy$-plane; (c) schematic of the magnetic topology following a symmetric pitchfork bifurcation in 2D, for comparison. When the field is 3D, $\partial B_{z}/\partial z \neq 0$ lending the field an open configuration, see text for details.}
\label{fig:model1}
\end{figure}

\begin{figure}
\centering
\includegraphics[width=0.5\textwidth]{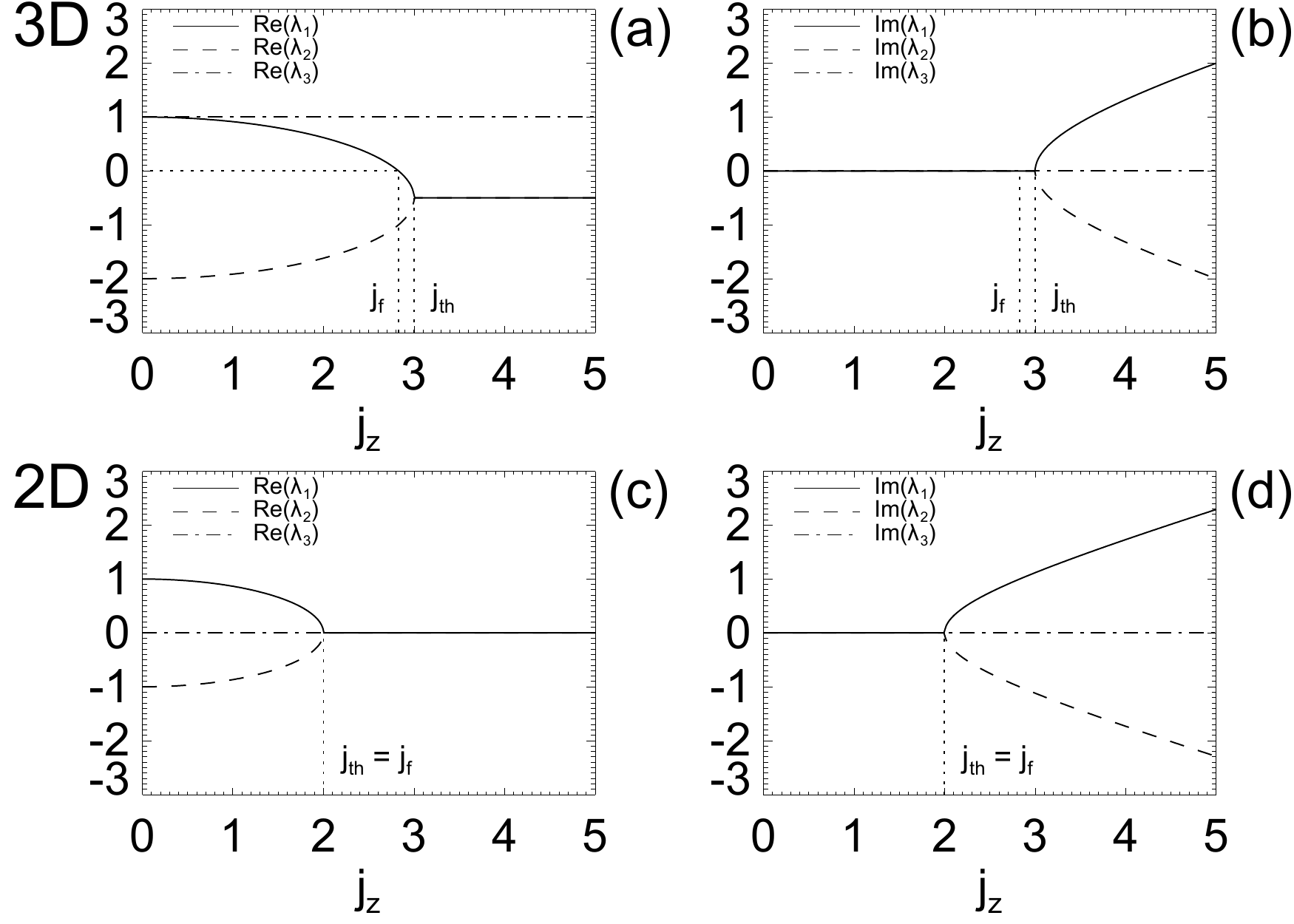}
\caption{Real and imaginary parts of the central null point/line eigenvalues as a function of current at the null ($j_{z}$) throughout the primary bifurcation in 2D ($\kappa=0$) and 3D ($\kappa=1$). Based on the model in Eq.~(\ref{primarybif}); in both cases $x_l=y_l=B_{0}=L_{0}=1$.}
\label{fig:evals}
\end{figure}

\subsection{Primary Bifurcations}
The original collapsed null (t.d.~$-1$) bifurcates via a pitchfork bifurcation \citep{Priest1996b} to form two nulls of t.d.~$-1$ flanking a spiral null of t.d.~$+1$, see Fig. \ref{fig:model1}a. This topology change is analogous to the formation of islands in 2D current sheets. However, there are several crucial differences. The first is that there exists no closed flux surface surrounding the spiral null (as about the O-point in 2D), but rather plasma and flux from both domains are efficiently mixed within the associated helical field structure, Fig.~\ref{fig:model1}b. This open, spiraling topology occurs because the field normal to the plane containing the three nulls ($B_{z}$) varies in the normal direction. That is, since $\partial B_z/\partial z\neq 0$, no closed field lines can exist in the $xy$-plane (as $\nabla \cdot {\bf B}=0$ within the volume requiring $\partial B_x/\partial x+\partial B_y/\partial y=-\partial B_z/\partial z\neq 0$). There is a further complication for our interpretation of the field structure and evolution introduced by the 3D topology shown in Fig. \ref{fig:model1}a. Typically flux ropes are thought of as twisted field regions with a strong ``guide field'' and no field reversals. Defining ``flux ropes'' as such, this bifurcation actually produces a pair of flux ropes with oppositely directed guide fields, so that each of the two spine lines of the spiral null forms the axis of one rope, and the fan of the spiral null lies on the interface of the two (against which these flux ropes splay out with a 3D stagnation-point geometry), Fig. \ref{fig:model1}a.

A simple model for the magnetic field that captures the essence of this initial bifurcation and demonstrates clearly the difference between the 2D and 3D pictures is given by:
\begin{equation}\label{primarybif}
\mathbf{B} = \frac{B_{0}}{L_{0}}[-(\kappa+1)x, y, \kappa z] + \boldsymbol{\nabla}\times \left( B_{0} j e^{-\frac{x^2}{{x_l}^2}-\frac{y^2}{{y_l}^2}} \hat{\mathbf{z}} \right),
\end{equation}
where $B_{0}$ and $L_{0}$ are some typical field strength and length scale. This field consists of a current cylinder, with strength and dimensions controlled by $j$ and $x_l$, $y_l$ respectively, added to a background linear null field centered on the origin. $\kappa=0$ sets the background to a 2D null line, and $\kappa=1$ produces a radial 3D null.

If $\kappa\neq 0$, as $j$ is increased the null point at the origin changes in nature from t.d.~$-1$ to t.d.~$+1$ at the point of bifurcation. An equivalent 2D measure, the Poincar\'{e} index \citep{Longcope2005,Molodenskii1977}, also exhibits a similar transition for the 2D null line ($\kappa=0$), which changes from $-1$ to $+1$ as $j$ is increased. The Jacobian evaluated at the origin is given by
\begin{equation}
\mathcal{M} = \frac{B_{0}}{L_{0}} \begin{pmatrix} -\kappa-1 & \frac{1}{2}(q-j_{z}) & 0\\ \frac{1}{2}(q+j_{z}) & 1 & 0\\ 0 & 0 &\kappa \end{pmatrix},
\end{equation}
where $q=2j(1/{x_l}^2-1/{y_l}^2)$ and $j_{z}=2j(1/{x_l}^2+1/{y_l}^2)$. The eigenvalues of this matrix can be written as 
\begin{eqnarray}
\lambda_{1,2} &=& -\frac{\kappa}{2}\pm \frac{1}{2}\sqrt{{j_f}^{2}+\kappa^2-j_{z}^2}, \\
\lambda_{3} &=& \kappa, 
\end{eqnarray}
where ${j_f}^{2} = j_{th}^2-\kappa^2$ and $j_{th}^{2}=(\kappa+2)^2+q^2$. Now, when $j_z$ is small we have $\lambda_{1,2} < 0$ and $\lambda_{3}>0$ corresponding to a null of t.d.~$-1$. If one now increases $j_z$, one reaches a critical threshold at $j_{z}^{2}={j_f}^{2}$ where the bifurcation occurs, and for $j_{z}^{2}>{j_f}^{2}$ we have  $\lambda_{2} < 0$ and $\lambda_{1,3}>0$ , i.e.~the null at the origin has changed to t.d.~$+1$. Increasing $j_z$ further, $\lambda_{1}$ and $\lambda_{2}$ become complex conjugates when $j_{z}^{2}>j_{th}^2$. Figure \ref{fig:evals} shows this transition in the eigenvalues as a function of $j_{z}$ for a 2D null line and a 3D null point. In the singular case of a 2D null line, ${j_f}^2=j_{th}^2$, therefore the transition in this case is directly from X-line (real eigenvalues) to O-line (purely imaginary eigenvalues). When $\kappa\neq 0$, $j_{th}^2 > {j_f}^2$ and the bifurcation initially creates a critical spiral with t.d.~$+1$ (a null where the fieldlines in the fan are aligned to two directions). Once $j_{z}^{2}>j_{th}^2$ the null becomes a regular spiral null as observed in the fields from our simulation -- see below. This simple model captures the change in the characteristics of the original 3D null point during the primary bifurcation process.

\begin{figure}
\centering
\includegraphics[width=0.5\textwidth]{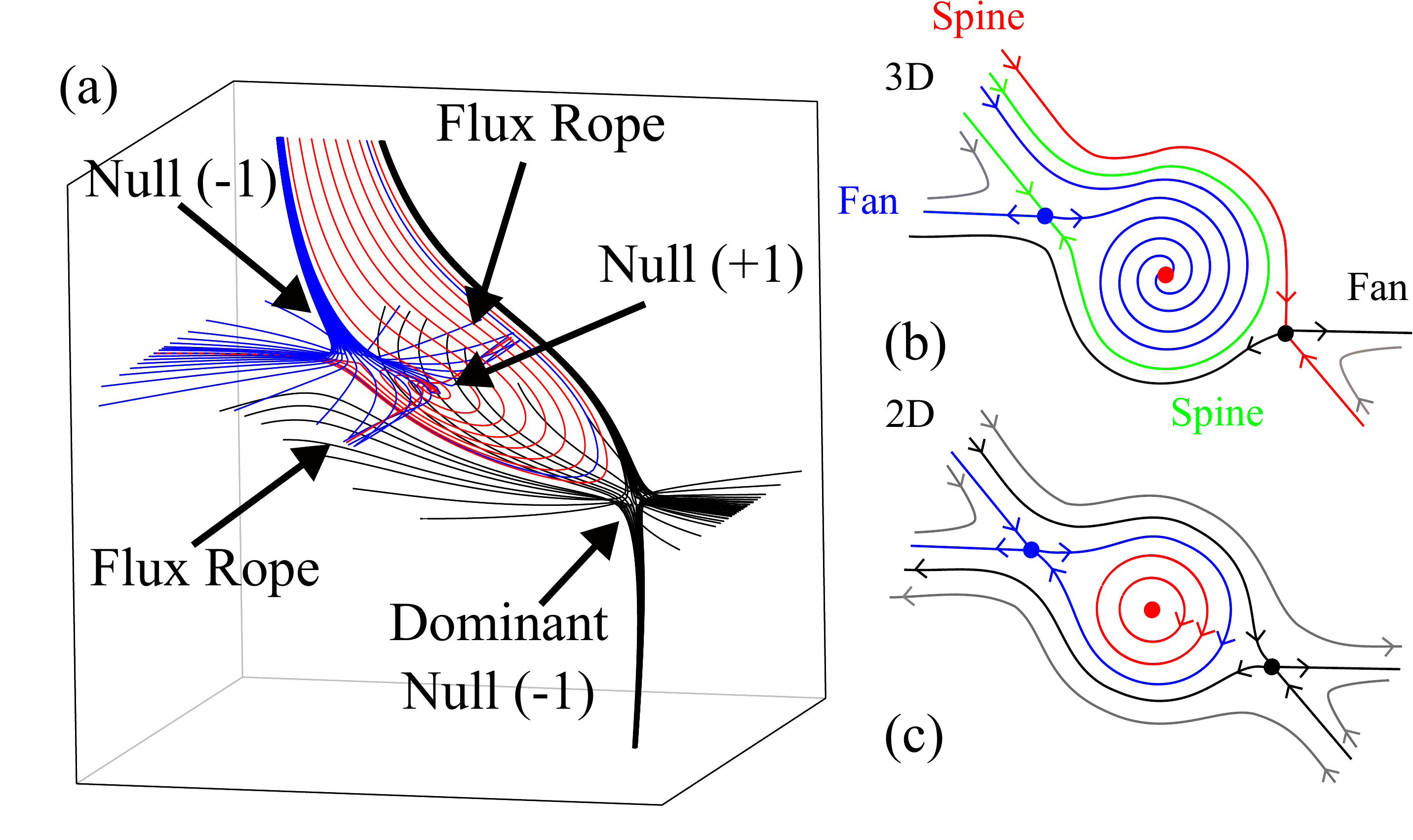}
\caption{(color online) Model of the field structure when the symmetry of the primary bifurcation is sufficiently broken. (a) the 3D field structure; (b) schematic of the field in the $xy$-plane; (c) schematic of the magnetic topology following an asymmetric pitchfork bifurcation in 2D for comparison. Generally, a 2D field immediately reverts to the topologically stable configuration shown in (c), whereas when the field is 3D a global spine-fan bifurcation is necessary as both the symmetric and asymmetric configurations are topologically stable, see text for details.}
\label{fig:model2}
\end{figure}

\begin{figure*}
\centering
\includegraphics[width=1.0\textwidth]{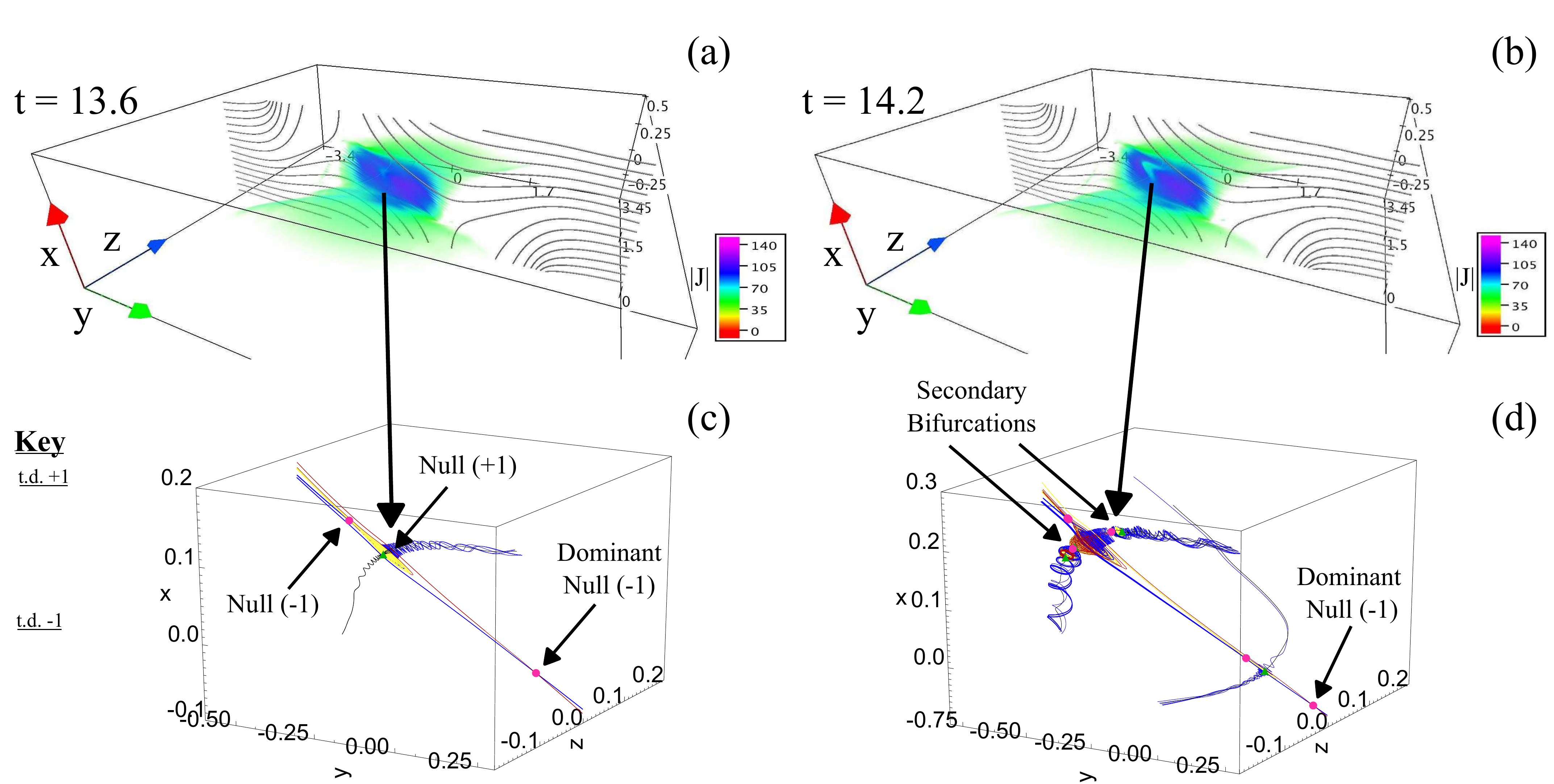}
\caption{(color online) The magnetic topology of the first flux rope pair formed following tearing within the current layer (Stage III). (a,b) show the overall shape of the magnetic field within the simulation as the flux ropes form. Grey fieldlines are plotted from line-tied points on the $y$-boundaries in the plane of spine-fan collapse ($z=0$). Shading indicates current density. (c,d) show the null point structure within the flux rope pair at each time. Pink circles indicate nulls with a t.d. of $-1$ and green triangles show nulls with a t.d. of $+1$. A small number of field lines are plotted to show the relative positions of the spines and fan planes of each null, colored according to the key on the left.}
\label{fig:bi1}
\end{figure*}

Once a bifurcation occurs, the value of $\kappa$ dictates the stability of the new magnetic configuration. When $\kappa=0$ (i.e.~when ${\bf B}$ is locally two-dimensional), the new field is topologically unstable since it contains two X-points connected by their separatrix lines, Fig. \ref{fig:model1}c. Any perturbation of this field will break this symmetry and the field will form the generic configuration shown in Fig. \ref{fig:model2}c. The four domains of the global field are now partitioned by the separatrices of a single X-point, which we refer to hereafter (following the usual convention) as the ``dominant X-point''. Since the configuration of Fig. \ref{fig:model1}c is inherently unstable, any evolving field will instantaneously revert to this second configuration.

By contrast, when the field is 3D ($\kappa\neq0$) the open topology of the central spiral means that a finite volume of flux (red field lines in Fig.~\ref{fig:model1}a, grey field lines in Fig.~\ref{fig:model1}b) separates the adjacent spines and fans of the two flanking 3D nulls. The greater the out-of-plane component of the field (the larger $\kappa$), the wider this corridor. Thus, this symmetric configuration -- in which a pair of separators connect the spiral null to each of the flanking nulls -- is topologically stable. This implies that the situation in which all three nulls lie on the separatrix surface separating the two topological domains will persist for a finite period of time following the pitchfork bifurcation (unlike the 2D case in which the generic case is to have a dominant null).

Nevertheless, for a sufficiently large perturbation of the system the symmetry of this null point triplet is eventually broken in the simulation -- as shown in {Fig.~\ref{fig:model2}b}. This occurs as one pair of nulls is caught in an outflow jet, and leaves the fan plane of a single null once more as the interface between the two global topological regions (referred to hereafter as the ``dominant null''). This requires a spine-fan bifurcation \cite{Brown1999}; a global topology change whereby the spine of one null crosses the fan of another, instantaneously becoming part of the fan in the process. As there is a finite envelope of flux to traverse for this bifurcation to occur, in our simulation this does not occur instantaneously (as it would in 2D), but rather after some finite time -- see Fig.~\ref{fig:bi1}a,c. Later in the simulation, once the field structure becomes more complex, spontaneous null pair creation occurs within the outflow region, leading directly to the latter configuration.

\begin{figure}
\centering
\includegraphics[width=0.4\textwidth]{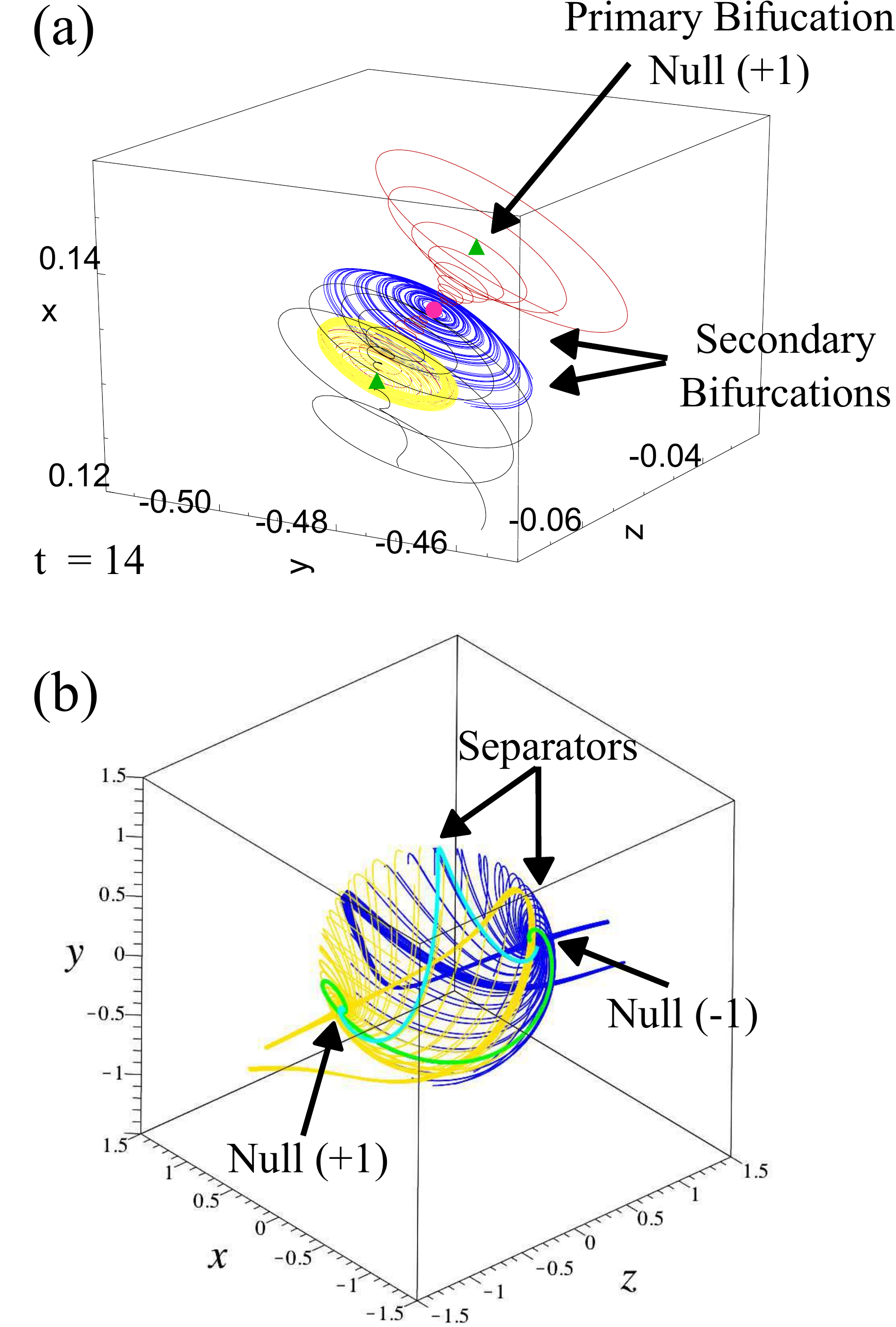}
\caption{(color online) (a) Close up view of the field topology of a secondary bifurcation at $t=14$ in the simulation. Spine and fan field lines are shown for the secondary nulls (following the key in Fig. \ref{fig:bi1}). The fan plane field lines have been truncated to better view the topology. (b) Analytical model of this secondary bifurcation. Yellow and blue field lines depict the magnetic field near the secondary nulls with t.d. $+1$ and $-1$, respectively. The fan surfaces intersect along two separators, shown in green and cyan. Model parameters: $B_{0}=-1, j=\phi_{0}=r_l=z_l=1$ and $k=0.2$. Note that in (a) the spine of the primary bifurcation null (t.d. $+1$) spirals around the entire secondary pair after helically wrapping the (red) spine of the nearest secondary null (t.d. $+1$), not shown.}
\label{fig:spheromaks}
\end{figure}

\subsection{Secondary Bifurcations}
Once the primary bifurcation creates the flux rope pair, this double rope structure is deformed by the strong reconnection outflow in the mid-plane, see Fig. \ref{fig:bi1}b,d. At this stage magnetic fluctuations lead to the formation of new null point pairs within the flux rope, in what we will refer to as secondary bifurcations. These new null pairs appear to form spontaneously near the axis of the flux rope, close to the spine of the spiral null (t.d. $+1$) produced by the primary bifurcation, Fig. \ref{fig:spheromaks}a. These spontaneous bifurcations lack a direct analogue in 2D, occurring as they do along the ``out of plane'' direction. However, they can be well described by models with cylindrical symmetry, to which some perturbation is added: making the field more generic.

Such a magnetic field that mimics the structure observed in the simulations following the bifurcations is given in cylindrical coordinates $(r,\phi,z)$ by 
\begin{equation}
\mathbf{B} = [0,r\phi_{0},B_{0}] + \boldsymbol{\nabla}\times \mathbf{A}_{1} + \mathbf{B}_{pert}, 
\end{equation}
where $\mathbf{A}_{1} = j r \exp\left(-\frac{r^2}{{r_l}^2}-\frac{z^2}{{z_l}^2}\right) \hat{\boldsymbol{\phi}}$. The field ${\bf B}_{pert}$ should be chosen to break the azimuthal symmetry to give a generic topology -- here we set $\mathbf{B}_{pert} = kz \hat{\mathbf{y}}$. The parameter $k$ controls the amplitude of this perturbation. For the case of exact symmetry ($k=0$) the two nulls of opposite degree are connected spine-to-spine and fan-to-fan in a spheromak configuration, see also \citet{Priest1996b}. This configuration is topologically unstable, and in the generic case ($k\neq 0$) where the symmetry is broken the fan planes intersect only along two separators \cite{hu2004}, Fig.~\ref{fig:spheromaks}b ($k=0.2$). As $k$ is increased from zero and the symmetry broken the isolated field within the spheromak becomes connected to the field outside. To be consistent with the outer magnetic field, the inner spines (those previously within the spheromak) become connected with magnetic field near the outer spines, Fig. \ref{fig:spheromaks}b. The fan surfaces of each of the two nulls become split into two regions bounded by the two separators; some fan field lines connect directly to external magnetic field, while some connect to field lines that lay inside the spheromak in the unperturbed field ($k=0$). The latter behave in a similar way to the inner spines, wrapping around on themselves before connecting with magnetic field originally outside the spheromak in the unperturbed field (as all of the previously isolated flux within the spheromak eventually must). However, as this is difficult to visualise the fan plane field lines in Fig. \ref{fig:spheromaks} have been truncated to give a clearer view of the spine lines which show agreement between the simulation field (black) and the model (yellow).

It is worth pointing out that such null pair formations are forbidden in an ideal evolution of the magnetic field. Thus, their presence within the flux ropes is a clear indicator that topology change is occurring within the flux ropes themselves, not just in the regions of high current density between them.

\begin{figure}
\centering
\includegraphics[width=0.4\textwidth]{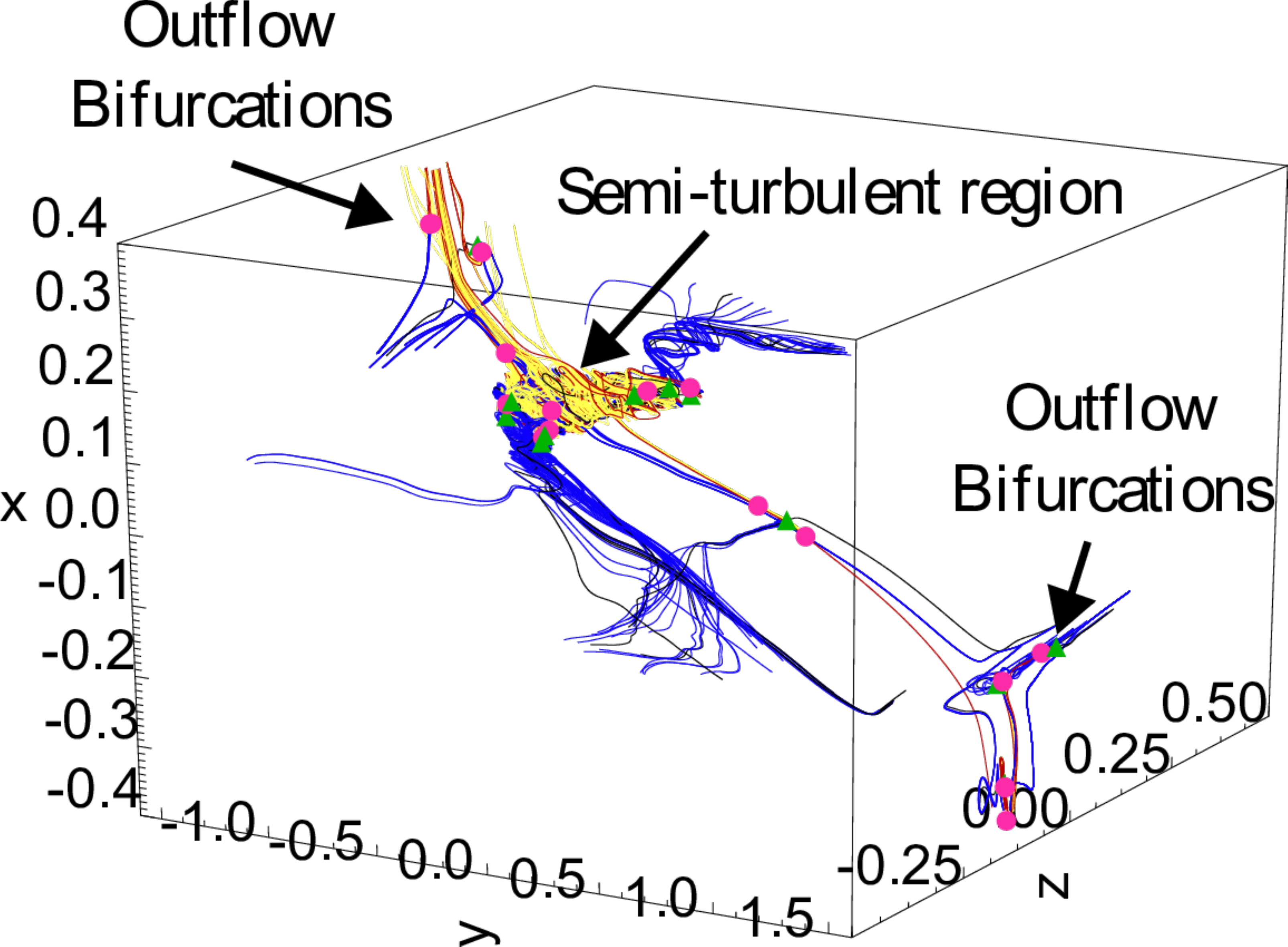}
\caption{(color online) A complex cluster of nulls in a turbulent-like region being advected towards the left outflow jet, with a new flux rope pair forming in its wake following a primary bifurcation ($t=21.5$). Field lines are traced from nearby each null point -- following the convention of Fig. \ref{fig:bi1}.}
\label{fig:cluster}
\end{figure}

\section{Turbulent-like Weak Field Evolution}
The magnetic topology becomes challenging to follow once the first flux rope pair is ejected and the system enters Stage IV, where newly formed flux ropes kink and begin to interact.  However, at least in the central region, the evolution described above generally follows. The flux rope pairs form via the primary bifurcation process described above. Prolific secondary bifurcations then occur within these flux rope structures, and near to the mid-plane ($z=0$) -- where the field in the current sheet is weakest -- clusters of nulls are formed as flux rope pairs begin to interact, discussed further below. Within these clusters null pairs are formed and annihilated rapidly. Figure \ref{fig:cluster} shows an example of a magnetic field with a cluster of nulls at the intersection of several interacting flux ropes on its way to being ejected, and a small flux rope pair beginning to form in its wake. The rapidly fluctuating and changing nature of the field within the null clusters may be the beginning of a turbulent evolution, but the lack of resolution within these regions prevents us from saying with any confidence that they exhibit true turbulence. Therefore, we refer to them simply as exhibiting a ``semi-turbulent'' or turbulent-like behavior, see also the discussion in Paper 1.

\begin{figure*}
\centering
\includegraphics[width=0.9\textwidth]{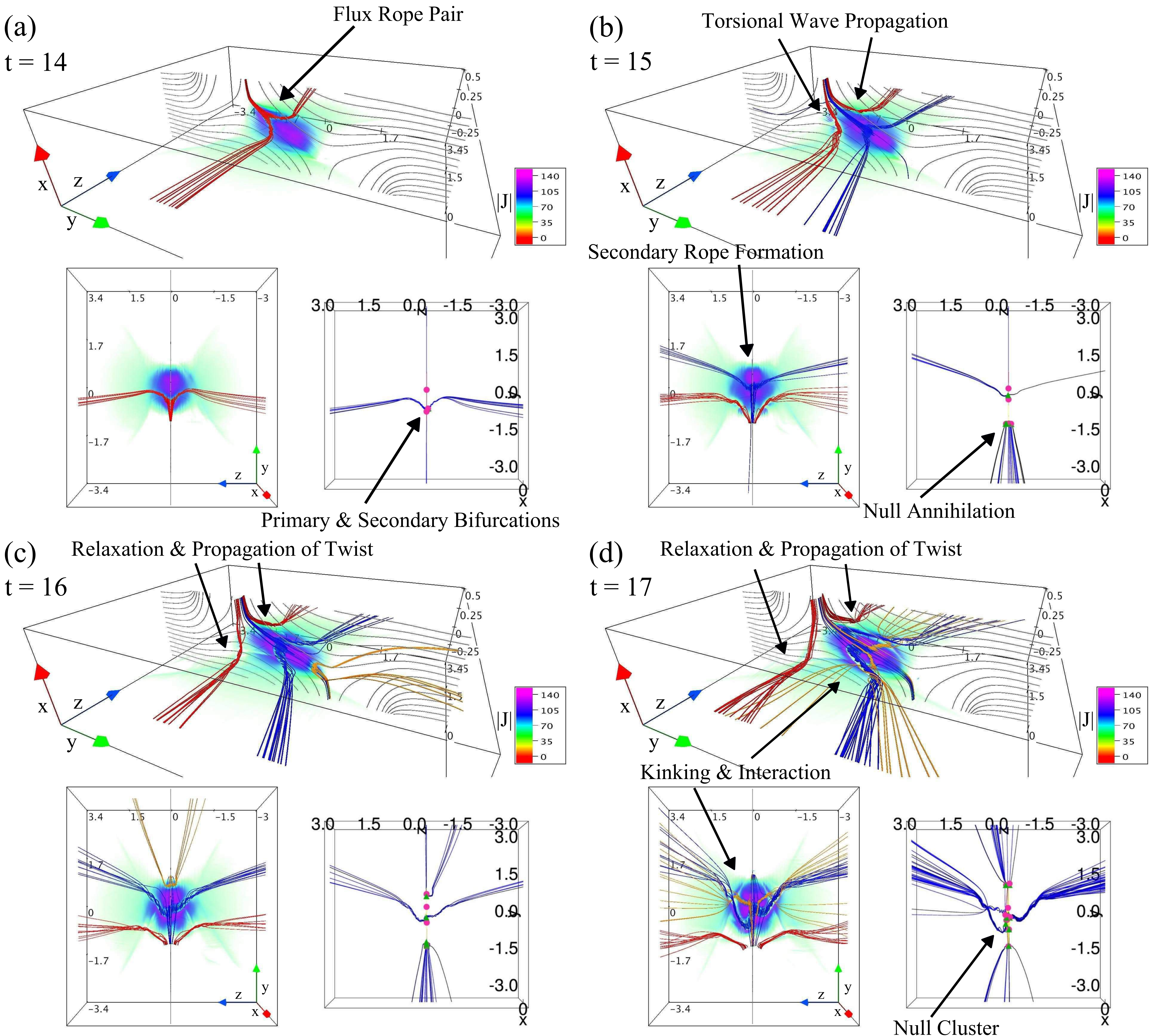}
\caption{(color online) Flux rope formation and ejection. Top and bottom-left panels: field lines within different flux rope pairs. The grey field lines are traced from fixed points on the $y=\pm 3.5$ boundaries, showing the field evolution in the mid-plane ($z=0$). Bottom right panels: field lines traced from near the null points. Pink circles show nulls with t.d. $-1$ and green triangles nulls with t.d. $+1$, see Fig. \ref{fig:bi1}. An animation of this figure is available online (Multimedia view).}
\label{fig:trip1}
\end{figure*}

\section{Outflow Jets}
The region where the outflow jet collides with ambient magnetic field is also highly complex with a large number of null points forming there. In the early stages the classical reverse current spike is seen to form \cite{Syrovatskii1971,Biskamp1986}. Soon after, this region becomes unstable to a shear flow instability and the outflow flails back and forth at regular intervals, generating turbulent eddies that sweep up the weak magnetic field in this region. It is not clear whether these eddies are the result of the Kelvin-Helmholtz instability, as described in the linear theory of \citet{Loureiro2013}, or perhaps the result of our asymmetric driving setup and line-tied external field configuration. The shear flow within these eddies also leads to the formation of short-lived magnetic null point pairs, Fig. \ref{fig:cluster}. Similar null generation in the outflow regions has recently been described by \citet{Karlicky2012} in 2D as a result of local enhancements in plasma-$\beta$ within sheared eddying outflows. We do not investigate these bifurcations further, but postulate that a similar process may be occurring. 

The regular formation of the outflow eddies is interrupted by the ejection of a flux rope pair, or multiple pairs connected via a null cluster. If the flux rope pair is small, as in the case of the first pair to be ejected (Fig. \ref{fig:bi1}), the central spiral null (t.d.~$+1$) of the rope catches up and annihilates with the collapsed null (t.d.~$-1$) ahead of it in the outflow region. The null pairs formed through secondary bifurcations within a given rope also then quickly annihilate with each other. When the combined structure is larger, as in the multi-rope example in Fig. \ref{fig:cluster}, the structure generates new null pairs as it slams into the static field in the outflow region (which is a region of relatively high plasma and magnetic pressure due to our closed boundary conditions and the finite extent of the imposed shear driving velocity). This is analogous to the null pair generation in the current layers that form between colliding islands in the fractal picture of 2D plasmoid accelerated reconnection \cite{Shibata2001}. However, in the present simulation this burst of additional nulls is short lived and all nulls in the structure also then quickly annihilate.

\begin{figure*}
\centering
\includegraphics[width=0.9\textwidth]{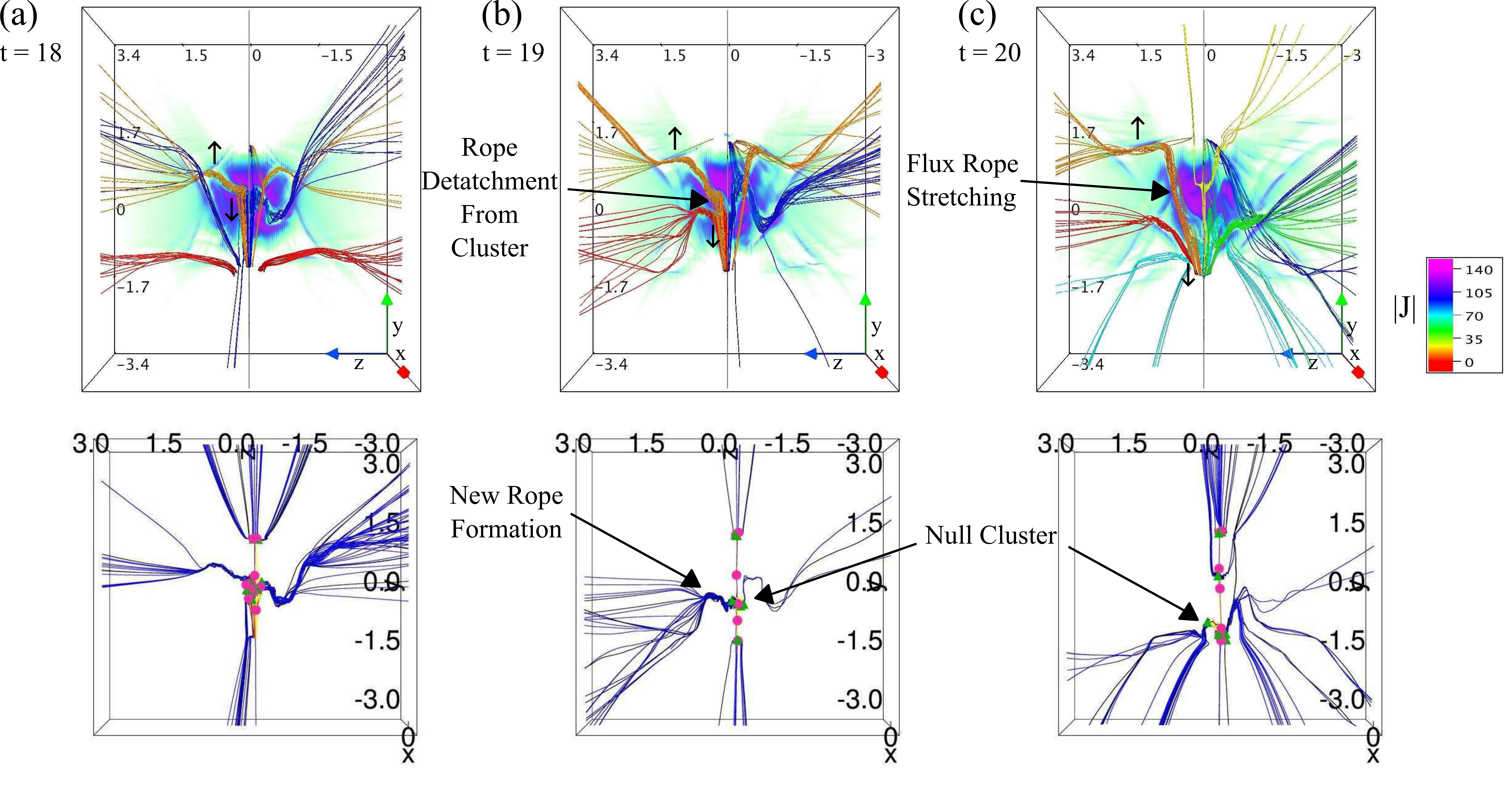}
\caption{(color online) Example of a flux rope (orange fieldlines) being stretched as different sections of the rope are caught in opposite outflow regions. Top panels -- field lines within selected flux ropes. Bottom panel -- positions of the magnetic nulls with field lines traced along their spines and fans as in Fig.~\ref{fig:bi1}.}
\label{fig:bridge1}
\end{figure*}

\section{Flux Rope Dynamics}
We now focus on the dynamics of the many flux ropes that form during the simulation. We emphasise again that such ``ropes'' are not distinct structures as might be envisaged by an O-line with an added guide field (see the earlier discussion). Rather the field within the ropes spirals inwards/outwards and is connected to the ambient field nearby which may have no twist, or may even be connected with another twisted ``rope'' structure. Therefore, our definition of a flux rope -- a region of helical, twisted field -- is somewhat arbitrary. However, these structures are co-located with channels of weak current in our fragmenting current layer, and are clearly important in controlling how reconnection proceeds in the layer. Furthermore, their dynamics can be complex, since they are susceptible to a 3D instability \cite{Dahlburg2001,Dahlburg2002} which kinks them so that they interact with one another and break up. We describe below their evolution and dynamics in the simulation.

\subsection{Formation, Propagation and Ejection}
As explained above, flux ropes can form in pairs -- connected by the spines of a spiral null. Once a pair of flux ropes has formed, the field near the spiral null in the mid-plane ($z=0$) is highly twisted compared with field further out along the spines of this null (the axis of each flux rope). As a consequence, bi-directional torsional MHD waves are launched along the ropes which allow the twist to propagate away from the centre of the current layer -- Fig. \ref{fig:trip1}b. These waves appear to travel close to the local Alfv\'{e}n speed, so we postulate that these are torsional Alfv\'{e}n waves. Note that these waves are launched in a direction that is nearly perpendicular to that of the outflow jets of the reconnection region. Similar three-dimensional spreading has been observed in laboratory experiments \cite{Katz2010} and two-fluid simulations \cite{Shepherd2012} of reconnection in setups with strong guide fields when reconnection is initiated in a localised region. As noted in Paper 1 this propagation of the twist away from the site of tearing, along with plasma ejection along the ropes, leads to a much flatter flux rope cross section than for islands in comparable 2D simulations. 

Depending on where a given flux rope pair forms relative to the large scale outflow jets of the main layer, different behaviors occur. If a pair forms near to an outflow jet, as in the example in Fig. \ref{fig:bi1}, then both ropes and the associated spiral null are ejected together in the outflow. The uni-directional outflow near the flux rope pair transports the whole structure, with the spiral null at the center advected fastest by the more rapidly outflowing plasma in the mid-plane. An example of this is shown by the red fieldlines in Fig. \ref{fig:trip1}a. As this occurs, the torsional MHD waves propagate outwards along each rope -- sweeping up the separatrix surface where the rope spans it and helping to reconnect flux across it, see Paper 1. The annihilation of the spiral null when it reaches the outflow is indicative of a disconnection between the two rope structures, Fig. \ref{fig:trip1}b. Both flux ropes then continue to be advected away from the current layer by plasma flow out of the mid-plane as the twist along their length spreads out and begins to relax, Fig. \ref{fig:trip1}c,d.

While some flux ropes are formed entirely within one outflow, others that form near the center of the layer can become highly stretched and distorted when different parts of the flux rope are caught in oppositely directed outflows. As such, the flux rope evolution becomes highly dynamic in Stage IV, once secondary kinking sets in. Figure \ref{fig:bridge1} -- orange fieldlines -- shows one such example. In this case, the section of the flux rope splayed out against the null cluster is advected downwards, whereas the rest of the rope is advected upwards towards the opposite outflow region, Fig. \ref{fig:bridge1}a. This stretches the rope as the two sections are moved further apart.

\begin{figure}
\centering
\includegraphics[width=0.47\textwidth]{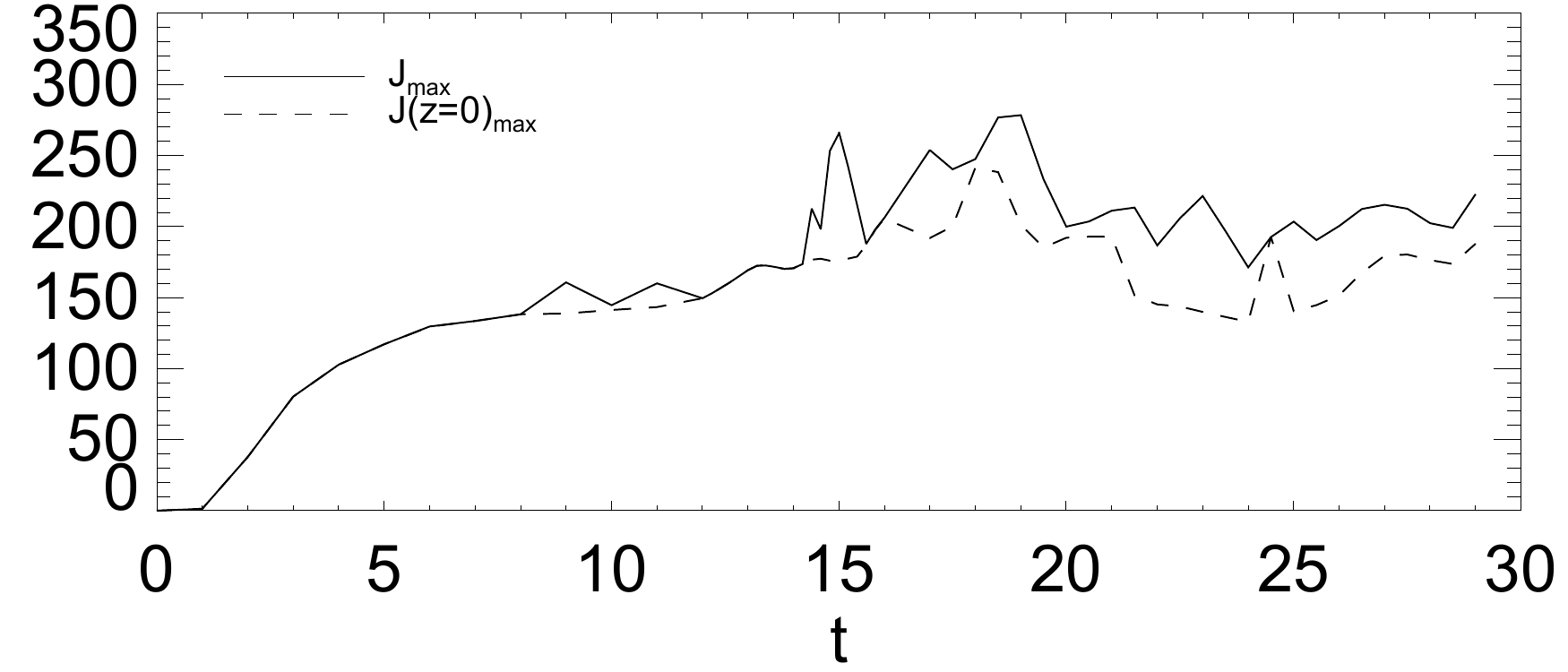}
\caption{The maximum current density in the volume compared with in the mid-plane ($z=0$). Note: the layer becomes unstable at $t\sim 10$.}
\label{fig:jmrr}
\end{figure}

Lastly, flux rope formation is not just limited to pair formation within the central region of the current sheet. This is particularly true once the main layer becomes highly fragmented. Flux ropes form when the tearing instability occurs over a finite patch of the current sheet, which subsequently spreads through the propagation of torsional MHD waves. In Stage III and early in Stage IV, these patches form in the mid-plane as by symmetry the strongest current occurs there. This generates the ``end on'' pairs of flux ropes discussed above, along with their associated null bifurcations. However, at later times (once these initial flux ropes become highly kinked) the current in the layer becomes patchy and fragmented. With the symmetry broken, the patches of highest current begin to be found out of the mid-plane, Fig. \ref{fig:jmrr}. Single flux ropes are then formed as these patches also begin to tear. However, no new nulls are generated as these ropes form away from the weak field of the mid-plane. 

\begin{figure}
\centering
\includegraphics[width=0.5\textwidth]{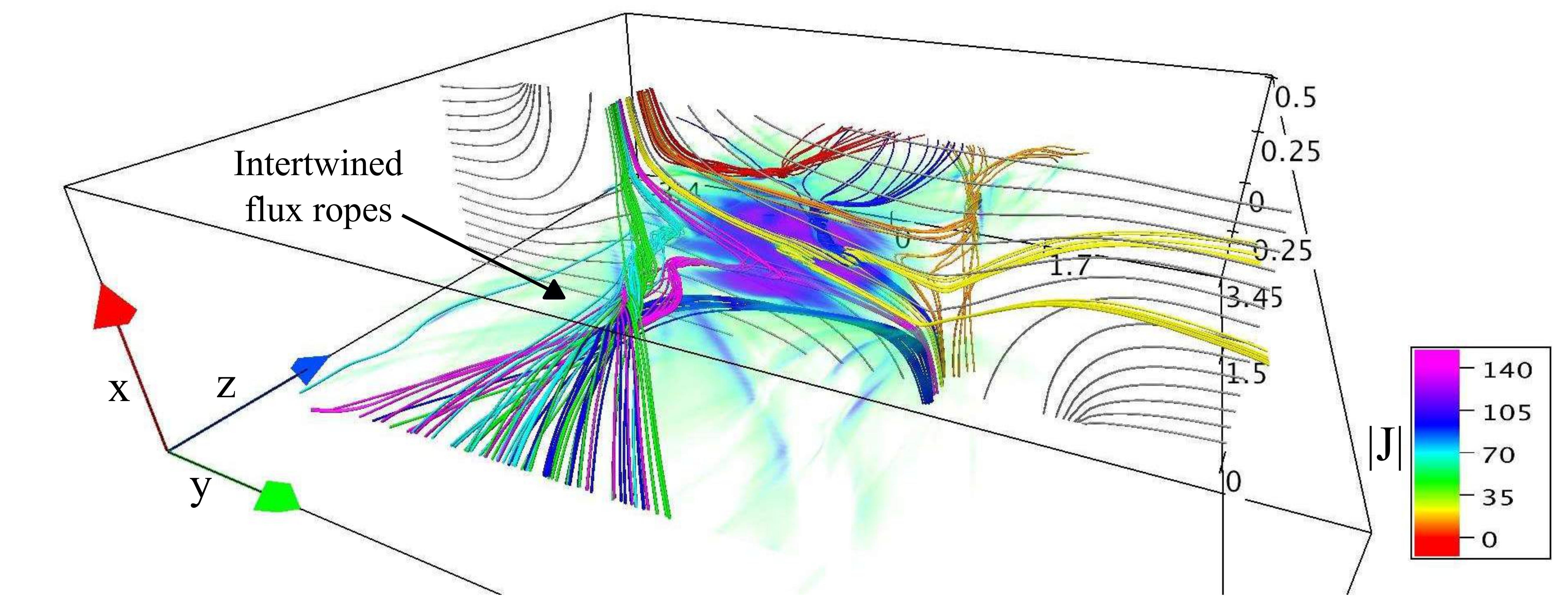}
\caption{(color online) Braiding of multiple flux ropes ($t=21$). Shown are field lines colored according to their individual flux rope.}
\label{fig:braid}
\end{figure}

\begin{figure}
\centering
\includegraphics[width=0.47\textwidth]{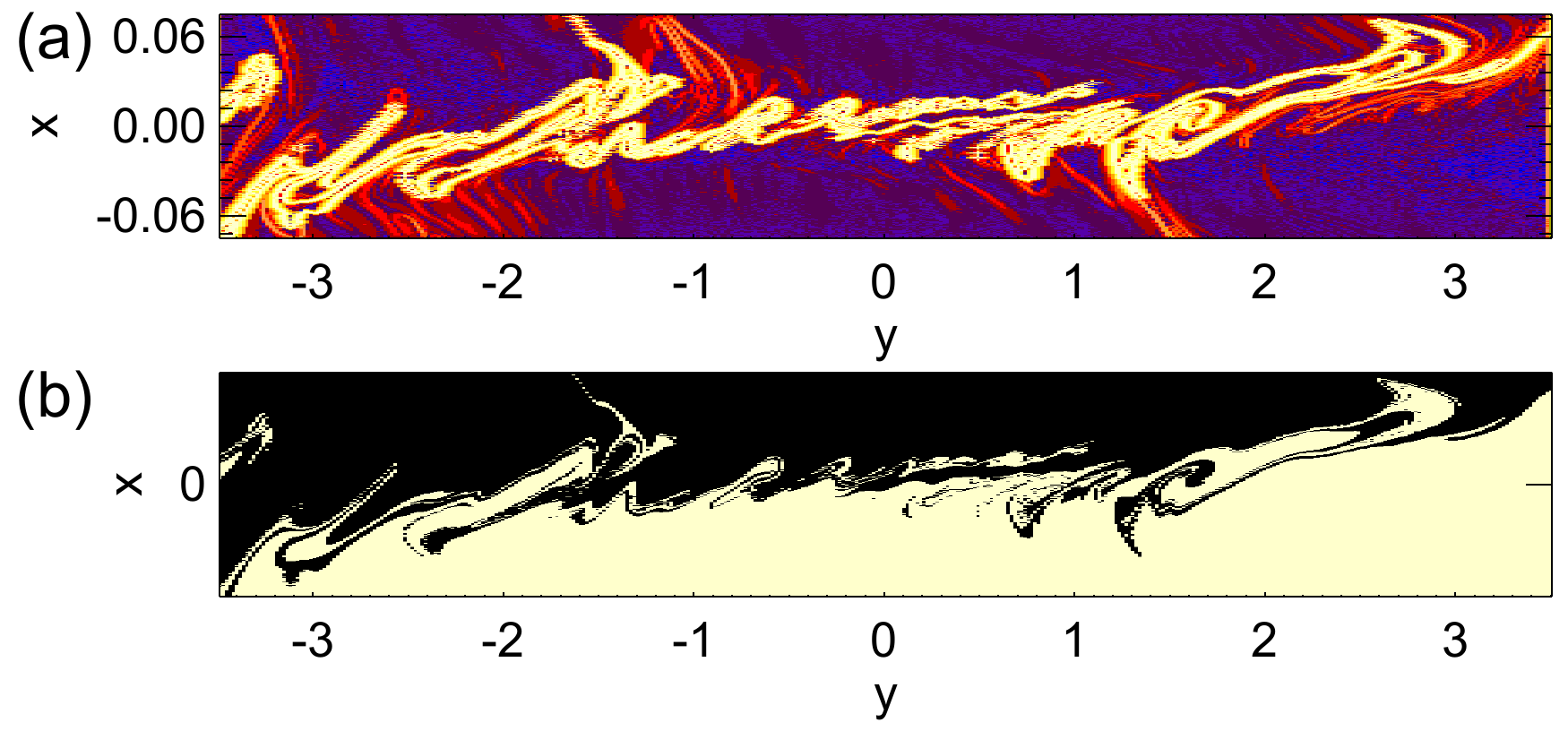}
\caption{(a) $\log{N}$ (capped at a finite value) showing the presence of QSLs nearby the separatrix surface. Produced at $t=21.5$. (b) Color map produced by plotting $80,000$ fieldlines from $z=-4.0$. Black indicates footpoints of field lines connected with the top of the box ($x=0.5$) and white those connected with the bottom of the box ($x=-0.5$). The separatrix surface lies at the intersection of the two domains.}
\label{fig:qsl}
\end{figure}

\subsection{Interaction}
A predominant feature of Stage IV in the evolution is the interaction of the flux ropes. Due to the direction of the magnetic shear outside the layer, the twist (or equivalently the sign of helicity \cite{Berger1999}) transferred to each of the flux ropes has the opposite handedness either side of the mid-plane ($z=0$) -- see Fig. \ref{fig:model1}. However, all flux ropes on the same side ($z \leq 0$, say) of the mid-plane have the same handedness of twist and so if two are brought into contact they merge into a larger rope structure. Figure \ref{fig:braid} shows an example of this --  four flux ropes (green, aqua, magenta and blue) have become wrapped into one another whilst propagating outwards from the mid-plane. Each started out as a localised twisted region but have been brought into contact following the onset of the ideal kinking instability -- see also the online material. This merging of twisted flux ropes in the outflow is a nice example of the upward cascade in scales of magnetic helicity that has recently been suggested by \citet{Antiochos2013} in their ``Helicity Condensation'' model to account for the smoothness of solar coronal magnetic fields at large scales (albeit on a much smaller scale in our case).

This continual formation near the center and ejection beyond the edges of the layer of relaxing flux rope structures generates a progressively more complex field in the vicinity of the separatrix surface as the simulation progresses. Newly forming ropes near the center thread into older relaxing ropes towards the edge of the sheet, which in turn thread into even older relaxing ropes beyond them. In this way, a complex layer of relaxing and propagating flux ropes is formed in the vicinity of the boundary between the two topological domains. 

This complexity is particularly evident in the footpoint mapping from the line-tied side boundaries of the simulation box. Figure \ref{fig:qsl}b shows a connectivity map from the $z=-4$ boundary coloured according to whether the footpoints connect with the top (black) or bottom (white) of the box. The flux ropes that straddle the separatrix surface generate the spirals in this color map and it is the evolution of these spirals that significantly enhances the flux transfer between the two topological regions, see Paper 1. However, not all flux ropes straddle the separatrix. This would be true of, for instance, the flux rope pair associated with the spiral null configuration depicted in Fig. \ref{fig:model2}a,b. The field within these spiral structures exhibits a large but finite change in the footpoint mapping and so should be visible as a Quasi-Separatrix Layer (QSL) \cite{PriestDemoulin1995}. A simple way of identifying field lines that pass through a QSL is by evaluating the norm of the Jacobian of the field line mapping:
\begin{equation}
N = \sqrt{\left(\frac{\partial Y}{\partial x}\right)^{2} + \left(\frac{\partial Y}{\partial y}\right)^{2} + \left(\frac{\partial Z}{\partial x}\right)^{2} + \left(\frac{\partial Z}{\partial y}\right)^{2}},
\end{equation}
where $(Y(x,y),Z(x,y))$ are the mapped coordinates on the top/bottom boundaries of the footpoints $(x,y)$ on the side boundary. Figure \ref{fig:qsl}a shows a contour plot of $\log{N}$. The discontinuity in the mapping across the separatrix surface shows up as the bright curve that follows the intersection of the black and white regions in Fig. \ref{fig:qsl}b. However, a number of QSL layers are evident away from this interface as ridges of high $N$, indicating that significant fine scale structure occurs not just on the separatrix, but also nearby.

\section{Discussion}

One of the more important conclusions of this work was to show that in addition to flux rope structures, spiral nulls are a major element of the magnetic topology when the tearing instability occurs in a weak, fully three-dimensional field. In particular, since flux ropes are found at all scales throughout the heliosphere the flux rope and the null point topology associated with it (especially the primary bifurcation) has implications in a number of areas. 

Starting with the smallest scales, kinetic simulation studies of tearing in 3D neutral sheets without guide fields have noted that spiral field structures form within the magnetic field following tearing \cite{Buchner1999,Wiegelmann2000}. As neutral sheets are topologically unstable, the tearing in these sheets must collapse to form a web of 3D magnetic nulls. Those nulls that are associated with flux ropes are likely to have the configurations associated with the primary or secondary bifurcations. 

This is also true of the magnetic configuration in the Earth's magnetotail -- which is often referred to as an X-line, but that must actually consist of many fluctuating 3D null points when the guide field is very weak or non-existent. Indeed, 3D spiral nulls have been identified from cluster data to exist within turbulence in the magnetotail \citep{Wendel2013,Xiao2006} and 3D kinetic simulations of tail reconnection \cite{Cai2006}. Our models help to explain the origins of these topological features. Elsewhere in the Earth's magnetosphere, the general field configuration of the polar cusps is one of a large-scale magnetic null \cite[e.g.][]{Stern1973}. Under northward IMF conditions reconnection occurs at a high aspect ratio current sheet formed over these regions. Global simulation and observational studies have noted the formation of flux ropes in the current sheets formed in these regions \cite[e.g.][]{Berchem1995,Le2001}. Our model for the primary bifurcation describes the formation and evolution of these flux ropes. Other simulations have observed that these 3D nulls bifurcate and form clusters \cite{Dorelli2007}. The subsequent dynamics described herein may also explain the formation of these multiple nulls. Of more general importance is the fact that we have shown that these flux ropes form in the vicinity of the separatrix surface, where they aid to drastically increase flux transport between the two topological regions (see Paper 1). Therefore, in this context flux rope formation may also help to mix the solar wind and magnetospheric plasma populations.

At even grander scales \citet{Masson2013} suggested a scenario based upon the breakout model \citep{Antiochos1999} to explain how impulsively accelerated Solar Energetic Particles (SEPs) are able to access open flux and escape into interplanetary space. Their 2.5D (2D with a constant or zero guide field) model relied upon the interchange of flux between domains divided by ``nulls'' at the intersection of closed flux surfaces. We have demonstrated (in agreement with previous works, e.g. \citet{lau1991,Daughton2011}) that such closed surfaces do not in general exist -- except in the case which they studied of a 2.5D field. The open flux rope structures formed in 3D are even more likely to aid in SEP transport, given the associated efficient mixing of flux between the two topological domains.

Concerning the general dynamics, we have also shown that the tearing instability is a natural mechanism for producing complex fields not just on the separatrix surface but also nearby it. This is in some ways similar to the S-web model \cite{Antiochos2011} proposed to explain the high latitudes at which the slow solar wind is observed. In this model a complex web of QSLs (resulting from deformations of the coronal hole boundary) surround the heliospheric neutral sheet, and are proposed as likely sites for reconnection. Similarly, the evolution of the flux ropes  in our simulations results in the creation of a series of QSLs in the near vicinity of the separatrix. However, the QSLs in the S-web model are associated with extra structure in the potential field, whereas by contrast our results demonstrate that this additional structure around the separatrix may be generated as a result of the reconnection process itself even in fields with much simpler global structure.

\section{Conclusions}

In this paper we examined the evolution of the magnetic field within the dynamic layer formed following the onset of the tearing instability in a current sheet generated about a 3D magnetic null point. The main motivation was to understand how topology change, flux rope formation and reconnection are linked in an evolving, tearing-unstable 3D null point current layer. Our main results can be summarised as follows:

\begin{enumerate}[(i)]
\item 
New null points are formed within the current layer in two main ways: (1) primary bifurcations -- analogous to island formation in 2D -- and (2) secondary bifurcations occurring within flux ropes, but without a direct 2D analogue. Both produce spiral nulls.
\item
By contrast with the 2D case, it is possible to have multiple nulls located on the global separatrix surface. A global topological (spine-fan) bifurcation is required when these nulls are ejected from the current layer to `detach' them from the global separatrix surface, leaving a single `dominant null'.
\item Flux ropes form in conjunction with null creation, but can also form independent of nulls, depending upon where the tearing occurs in the current layer.
\item Flux rope interaction continually increases the complexity of the magnetic field in the vicinity of the separatrix, broadening the overall width of the non-ideal layer.
\item Localised tearing is a source of torsional MHD waves, launched at an angle to the main reconnection outflow jets.
\end{enumerate}

\begin{acknowledgements}
Financial support from the Leverhulme Trust and fruitful discussions with K. Galsgaard, G. Hornig, A. Haynes and A. Russell are gratefully acknowledged. DP also acknowledges financial support from the UKÕs STFC (grant number ST/K000993). Computations were carried out on the UKMHD consortium cluster funded by STFC and SRIF.
\end{acknowledgements}

%

\end{document}